\documentclass{optica-article}

\journal{oe}

\articletype{Research Article}

\usepackage{lineno}
\usepackage{hyperref}
\usepackage{url}

\begin{document}
 
\title{Bayesian optimization of Bose-Einstein condensation via evaporative cooling model}

\author{Jihao Ma,\authormark{1,2$\dagger$} Ruihuan Fang,\authormark{1,2$\dagger$} Chengyin Han,\authormark{3,1} Xunda Jiang,\authormark{1,2,4} Yuxiang Qiu,\authormark{1,2} Zhu Ma,\authormark{1,2} Jiatao Wu,\authormark{1,2} Chang Zhan,\authormark{1,2} Maojie Li,\authormark{1,2} Bo Lu,\authormark{3,1*} and Chaohong Lee\authormark{3,1,2**}}

\address{\authormark{1}Guangdong Provincial Key Laboratory of Quantum Metrology and Sensing \& School of Physics and Astronomy, Sun Yat-Sen University (Zhuhai Campus), Zhuhai 519082, China\\
\authormark{2}State Key Laboratory of Optoelectronic Materials and Technologies, Sun Yat-Sen University (Guangzhou Campus), Guangzhou 510275, China\\
\authormark{3}College of Physics and Optoelectronic Engineering, Shenzhen University, Shenzhen 518060, China\\
\authormark{4}School of Physics and Optoelectronic Engineering, Foshan University, Foshan 528000, China\\
\authormark{$\dagger$}These authors contributed equally to this work}

\email{\authormark{*}lubo1982@szu.edu.cn} 
\email{\authormark{**}chleecn@szu.edu.cn} 

\begin{abstract}
    To achieve Bose-Einstein condensation, one may implement evaporative cooling by dynamically regulating the power of laser beams forming the optical dipole trap.
	We propose and experimentally demonstrate a protocol of Bayesian optimization of Bose-Einstein condensation via the evaporative cooling model.
	Applying this protocol, pure Bose-Einstein condensate of $^{87}$Rb with $2.4\times10^4$ atoms can be produced via evaporative cooling from the initial stage when the number of atoms is $6.0\times10^5$ at a temperature of 12 $\mu$K.
	In comparison with Bayesian optimization via blackbox experiment, our protocol only needs a few experiments required to verify some close-to-optimal curves for optical dipole trap laser powers, therefore it greatly saves experimental resources.
\end{abstract}

\section{Introduction}
Evaporative cooling, a powerful technique for preparing Bose-Einstein condensate (BEC) \cite{1,PhysRevLett.75.3969,PhysRevLett.75.1687}, is usually optimized via dynamically regulating the power of laser beams forming the optical dipole trap.
The main goal of optimizing evaporative cooling is to increase the number of atoms that reach Bose-Einstein condensation.
Several Bayesian optimizations\cite{7352306} based on Gaussian process regression (GPR) have been applied for optimizing the creation of BEC\cite{4,Nakamura:19,PhysRevA.102.011302,Barker_2020,PhysRevResearch.4.043216}.	
GPR establishes a statistical model of the relationship between the controllable input parameters characterizing the power of laser beams and the experimental output of the quality of the BEC, thus it is able to statistically predict the close-to-optimal input parameters.
It assumes that no prior knowledge of evaporative cooling can be obtained, thus it is suitable for optimizing complex evaporative cooling processes.
However, it requires a large run number of experiments to search the close-to-optimal input parameters, which would consume a lot of experimental time and resources.
On the other hand, one may optimize evaporative cooling via an experiment-independent model based on classical kinetic theory\cite{PhysRevA.53.381}, which approximately describes the process of evaporative cooling, but it becomes invalid when the system approaches the critical point of BEC phase transition.

In this paper, we propose a protocol of Bayesian optimization of Bose-Einstein condensation via the evaporative cooling model (ECM) and demonstrate it experimentally.
In our protocol, the blackbox experiments (BBE) of evaporative cooling are replaced by numerical simulation of the ECM, and Bayesian optimization is used for searching close-to-optimal curves of the optical trap power.
Among the close-to-optimal curves attained by our protocol, we experimentally demonstrate that pure BEC of $^{87}$Rb with $2.4\times10^4$ atoms can be prepared via evaporative cooling from the initial stage when the number of atoms is $6.0\times10^5$ at temperature 12 $\mu$K.
Our protocol does not depend on experimental data in the process of optimization, but only a few experiments are required to verify some close-to-optimal curves of optical trap power. In comparison with Bayesian optimization based on the BBE in which real-time experimental data is required as feedback in each cycle, our protocol can greatly save experimental resources.

\section{Bayesian optimization based on evaporative cooling model}	

\subsection{Evaporative cooling model}

We consider the evaporative cooling implemented in a crossed optical dipole trap (ODT), which is formed by two focused Gaussian laser beams with powers $P_L$ and $P_R$.
Both laser beams propagate in the $x-y$ plane, and their foci overlap at the origin of the coordinate.
In addition to the optical potential, the crossed ODT includes gravity along the $z$-axis.
Thus the whole potential can be given as $U=-B\sum_{k=L,R} \frac{2P_k}{\pi w_{k}^2(x_k)}exp\left[{-2\frac{y_k^2+z^2}{w_{k}^2(x_k)}}\right]+mgz$, where $B=-\frac{3\pi c^{2}}{2\omega_{\text {eff}}^{3}} \left(\frac{\Gamma_{\text {eff}}}{\omega_{\text{eff }}-\omega}+\frac{\Gamma_{\text{eff}}}{\omega_{\text{eff}}+\omega}\right)$ with $\omega_{\text {eff}}=\frac{1}{3} \omega_{D1}+\frac{2}{3} \omega_{D2}$ and $\Gamma_{\text{eff}}=\frac{1}{3}\Gamma_{D1}+\frac{2}{3}\Gamma_{D2}$ being the effective transition and the effective line width defined by the weighted average of both the $D1$ and $D2$ lines of $^{87}$Rb atoms\cite{GRIMM200095,Xiong:10},  $x_k=y\sin\phi_k+x\cos\phi_k$ and $y_k=y\cos\phi_k-x\sin\phi_k$ ($\phi_k$ being the angle between the propagating direction of the laser beam and $x$-axis), $m$ is the mass of atom, and $g$ is the gravitational acceleration.
The beam size $w_{k}(x_k)=w_{k,0}\sqrt{1+(\frac{x_k}{x_{k,R}})^2}$ is determined by the beam waist $w_{k,0}$, the Rayleigh range $x_{k,R}={\pi w_{k,0}^2}/{\lambda}$ and the laser wavelength $\lambda$.  
Two key parameters for evaporative cooling are the trap depth $\epsilon_t$ and the mean trap frequency $\bar{\omega}=\left(\omega_{x} \omega_{y} \omega_{z}\right)^{1 / 3}$, where $\omega_{x}$, $\omega_{y}$ and $\omega_{z}$ represent trap frequencies along $x$-axis, $y$-axis, and $z$-axis, respectively.
Because the ODT potential is tilted by the gravitational potential, the trap depth is calculated as $\epsilon_t=\left|U(z_1)-U(z_2)\right|$, here $z_1$ and $z_2$ are two points that the derivative of trap potential $\frac{\partial}{\partial z} U(x=0,y=0,z)$ equal to zero.
The trap potential $U$, the trap depth $\epsilon_t$, and mean trap frequency $\bar{\omega}$ are also related to the trap laser power. 

Evaporative cooling includes many complicated processes.
Here, we consider the effect of evaporation, one-body loss due to collisions with background gas, three-body loss, and the losses due to the trap changes. The other effects are ignored.
According to the kinetic theory\cite{PhysRevA.53.381}, the total atom number $N$ and the atom temperature $T$ during evaporation in a three-dimensional harmonic trap (as an approximation of crossed ODT) obey \cite{PhysRevA.87.053613, PhysRevA.93.043403}
\begin{equation}
  \dot{N}=-\left[\Gamma_{ev}(N,T,\epsilon_{t},\bar{\omega})+\Gamma_{3 B}(N,T,\bar{\omega})+\Gamma_{1 B}\right] N,
  \label{number of atoms}
\end{equation}
\begin{equation}
  \dot{T}=-\left[\frac{\Gamma_{ev}(N,T,\epsilon_{t},\bar{\omega})}{3}{\widetilde{\kappa}(\epsilon_{t})} -\frac{\Gamma_{3 B}(N,T,\bar{\omega})}{3}-\frac{\dot{\bar{\omega}}}{\bar{\omega}}-1\right] T.
  \label{temperature}
\end{equation}
where $\Gamma_{e v}(N,T,\epsilon_{t},\bar{\omega})=[\eta-4\mathcal{P}(4, \eta) / \mathcal{P}(3, \eta)] e^{-\eta} n_0 \sigma \bar{v}$ is the evaporation rate \cite{PhysRevA.104.033313}, 
in which $\eta=\epsilon_{t}/k_{B} T$ with Boltzmann's constant $k_{B}$,
$n_0=N \bar{\omega}^{3}\left[m/(2\pi k_{B}T)\right]^{3/2}$ is the atom peak density,
$\sigma=8 \pi a_{s}^{2}$ is the elastic cross section for identical bosons with $s$-wave scattering length $a_{s}$$=$$98 a_{0}$ for $^{87}$Rb (with $a_{0}$ being Bohr radius) \cite{PhysRevLett.88.093201},
$\bar{v}=\left(8 k_{B} T / \pi m\right)^{1 / 2}$  is the average velocity of the atoms,
and $\mathcal{P}(a, \eta)$ is the incomplete gamma function\cite{PhysRevA.53.381}.
The three-body loss rate $\Gamma_{3 B}(N,T,\bar{\omega})=K_{3 B} n_{0}^{2} /(3 \sqrt{3})$ with $K_{3 B}=4.3(\pm 1.8) \times 10^{-29} \mathrm{~cm}^{6} / \mathrm{s}$ is a temperature-independent three-body inelastic loss rate coefficient for $^{87}$Rb in the  F=1  ground state\cite{PhysRevA.87.053613,PhysRevLett.79.337}. 
The average energy for escaped atom is $\widetilde{\kappa} k_{B} T$ with $\widetilde{\kappa}=\eta+ 1-\frac{\mathcal{P}(5, \eta)}{\eta \mathcal{P}(3, \eta)-4\mathcal{P}(4, \eta)}$.
$\Gamma_{1 B}=1 / 128 \mathrm{~s}^{-1}$ is the rate for one-body loss determined by background collisions in our experiment.

We call Eqs.(\ref{number of atoms}-\ref{temperature}) for the evolution of total atom number and atom temperature as the ECM.
As the change of laser beam powers $P_L$ and $P_R$ with time is known, given the initial number $N_0$ and initial temperature $T_0$, the evolution of total atom number $N$ and atom temperature $T$ during evaporative cooling can be obtained via numerical calculation of the ECM.
Therefore, one could change the power of laser beams $P_L$ and $P_R$ for optimizing the evaporative cooling process to obtain large number of atom reaching BEC.

\subsection{Bayesian Optimization}
\begin{figure}[htbp]
  \centering\includegraphics[width=10cm]{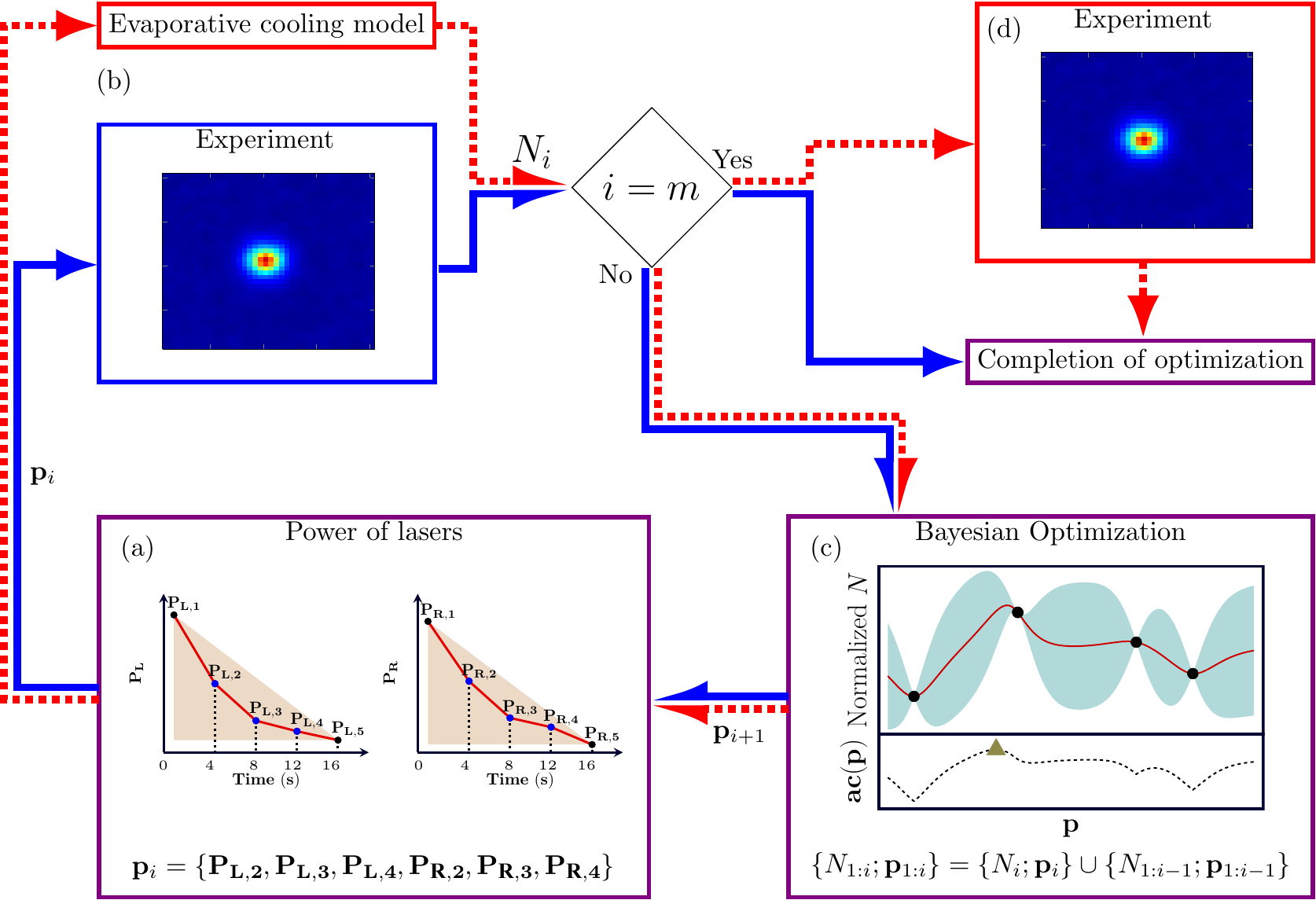}
  \caption{
  The Bayesian optimization processes for creating BEC based on the ECM scheme (red dashed arrows) and BBE scheme (blue solid arrows).  
  (a) The red polylines indicate the change of ODT laser powers. The black dots ($P_{L,1}$, $P_{L,5}$) and ($P_{R,1}$, $P_{R,5}$) represent the startpoint and endpoint of each beam power. The blue dots $P_{L,2-4}$ and $P_{R,2-4}$ represent the turning points of each beam power. The brown areas represent the bound of each beam power. $\mathbf{p}_i$ is the vector of intermediate ODT laser powers.
  (b) Both the ECM scheme and BBE scheme obtain an output atom number $N_i$ via the input $\mathbf{p}_i$. The atom number $N_i$ based on the BBE scheme is obtained by absorption imaging. 
  (c) The atom number $N_i$ and corresponding input $\mathbf{p}_{i}$ are merged into data set $N_{1:i-1}$ and $\mathbf{p}_{1:i-1}$. 
  The Bayesian optimization algorithm fits the data set $\{N_{1:i};\mathbf{p}_{1:i}\}$ (black circle dots) using GPR.
  The red solid line and teal area represent the mean of GPR and its standard variance, and the black dashed line is the acquisition function.
  The next input $\mathbf{p}_{i+1}$ is obtained by searching the maximum of the acquisition function.
  The optimization repeats until the number of loops $i$ reaches a selected number $m$.
  (d) The process based on the ECM will apply the inputs with optimal parameters to the experiment.
  }
  \label{figure1}
\end{figure}

To find out the optimal curves of ODT beam powers for obtaining a large atom number BEC, the Bayesian optimization based on GPR is used\cite{snoek_practical}.
We split the evaporative cooling process into four stages with equal time durations and change the intermediate power of the ODT beams $\mathbf{p}_i=\left\{P_{L,2},P_{L,3},P_{L,4},P_{R,2},P_{R,3},P_{R,4}\right\}$ of the $i$-th loop \cite{Nakamura:19}, as shown in Fig. \ref{figure1}(a). 
The brown shadows in Fig.~\ref{figure1}(a) represent the constraint of ODT beam powers. 
Then the corresponding atom number $N_i$ can be obtained through either the ECM or experiments.
The atom number $N_i$ and corresponding input $\mathbf{p}_{i}$ are merged into data set $N_{1:i-1}$ and $\mathbf{p}_{1:i-1}$.
The Bayesian optimization based on GPR is used to fit a statistical model of the relationship between the set of input parameter $\mathbf{p}_{1:i}$ and the set of output atom number $N_{1:i}$, as shown in Fig.~\ref{figure1}(c).
We normalize the target value $N_{1:i}$ as $N\left(\mathbf{p}_{1:i}\right)$ by removing the mean and scaling to unit variance,
\begin{equation}
  N\left(\mathbf{p}_{1:i}\right) = \left(N_{1:i} - \bar{N}_{1:i}\right) / \mathrm{std}(N_{1:i}),
\end{equation}
where $\bar{N}_{1:i}$ and $\mathrm{std}(N_{1:i})$ are the mean and statistic standard variance of obtained atom numbers $N_{1:i}$.
With the GPR model, we get the conditional distribution of unmeasured input parameter set $\mathbf{p}_{1:j}^{*}$ under the $\mathbf{p}_{1:i}$.
Considering that $N(\mathbf{p}_{1:i})$ and $N(\mathbf{p}_{1:j}^{*})$ obey the multivariate normal distribution
\begin{equation}
  \begin{bmatrix}
    N\left(\mathbf{p}_{1:i}\right)\\
    N(\mathbf{p}^{*}_{1:j})
  \end{bmatrix}
  \sim
  \mathcal{N}\left(
      0,
      {\Sigma}
  \right),
\end{equation}
where the covariance matrix is
\begin{equation}
  \mathbf{\Sigma}=\begin{bmatrix}
    K(\mathbf{p}_{1:i}, \mathbf{p}_{1:i})+\sigma_i^2 I & K(\mathbf{p}_{1:i}, \mathbf{p}^{*}_{1:j})\\
    K(\mathbf{p}^{*}_{1:j}, \mathbf{p}_{1:i}) & K(\mathbf{p}^{*}_{1:j}, \mathbf{p}^{*}_{1:j})
  \end{bmatrix},
\end{equation}
and $K$ denotes the covariances evaluated at all pairs of training and test points, $\sigma_i^2=10^{-6}$ is the noise level.
To describe the covariance between points of parameter space, we use the Mat\'ern class of covariance functions as the Kernel function of 
$K(\mathbf{p}_i, \mathbf{p}_j)=\frac{2^{1-\nu}}{\Gamma(\nu)}\left(\frac{\sqrt{2\nu}r}{\ell}\right)^{\nu}K_\nu\left(\frac{\sqrt{2\nu}r}{\ell}\right)$,
with positive parameters $\nu$ and $\ell$, the distance of input space $r=|\mathbf{p}_i-\mathbf{p}_j|^2$, and $K_\nu$ is a modified Bessel function. 
In our case, we select $\nu=2.5$ to avoid evaluating the modified Bessel function.
We can get the conditional distribution of $N(\mathbf{p}_{j}^*)$
\begin{equation}
  N(\mathbf{p}_{j}^*)|\mathbf{p}_{1:i},N(\mathbf{p}_{1:i}),\mathbf{p}_{j}^*\sim \mathcal{N}(\bar{N}(\mathbf{p}_{j}^*), N_{\mathrm{std}}(\mathbf{p}_{j}^*)),
\end{equation}
with expected value
$\bar{N}(\mathbf{p}_{j}^*)=K(\mathbf{p}_{j}^*, \mathbf{p}_{1:i})[K(\mathbf{p}_{1:i},\mathbf{p}_{1:i})+\sigma_i^2 I]^{-1}N(\mathbf{p}_{1:i}),$
and standard variance
$N_{\mathrm{std}}(\mathbf{p}_{j}^*)=K(\mathbf{p}_{j}^*,\mathbf{p}_{j}^*)-K(\mathbf{p}_{j}^*,\mathbf{p}_{1:i})[K(\mathbf{p}_{1:i},\mathbf{p}_{1:i})+\sigma_i^2 I]^{-1}K(\mathbf{p}_{1:i},\mathbf{p}_{j}^*).$
The super parameters $\ell$ is initially select as $1$ and will be optimized through maximizing the log-marginal likelihood $\log p\left[N(\mathbf{p}_{1:i})|\mathbf{p}_{1:i}\right]$ for every 5 GPR updating,
\begin{equation}
  \begin{aligned}
  \log p\left[N(\mathbf{p}_{1:i})|\mathbf{p}_{1:i}\right] =& -\frac{1}{2}N(\mathbf{p}_{1:i})^{\mathcal{T}}[K(\mathbf{p}_{1:i},\mathbf{p}_{1:i})+\sigma_i^2 I]^{-1} N(\mathbf{p}_{1:i}) \\
  &- \frac{1}{2} \log |K(\mathbf{p}_{1:i},\mathbf{p}_{1:i})+\sigma_i^2 I| - \frac{i}{2} \log 2\pi.
  \end{aligned}
\end{equation}

The predicted parameter $\mathbf{p}_{i+1}$ is obtained via maximizing the acquisition function.
The acquisition function for Bayesian optimization can be obtained by the surrogate model. Using the Upper Confidence Bound, the acquisition is expressed as 
\begin{equation}
\mathrm{ac}(\mathbf{p})=\bar{N}(\mathbf{p}) + \kappa N_{\mathrm{std}}(\mathbf{p}),
\end{equation}
with $\kappa=1$ in our case, and thus the next input $\mathbf{p}_{i+1}=\arg\max_{\mathbf{p}} \mathrm{ac}(\mathbf{p})$.
The optimization will be terminated until the loops number of optimization $i$ reaches a selected target $m$.

The optimization algorithm is based on the Bayesian optimization package\cite{Fernando}.
After the optimization, some close-to-optimal curves obtained via Bayesian optimization based on the ECM will be applied in the experiment to verify the result, as shown in Fig. \ref{figure1}(d).

\section{Experimental demonstration}
\subsection{Experimental setup}
The experimental apparatus that we use is described in more detail in \cite{Zhu_Ma_103701}. It mainly consists of two glass chambers with antireflection coating from 700 nm to 1100 nm.
The vacuum pressures of the first chamber and the second chamber are about $1\times 10^{-7}$ Pa and $3\times 10^{-9}$ Pa, respectively.
Atoms are collected in the first chamber and pushed to the second chamber within 5 s.
After the Magnetic Optical Trap (MOT) process, we compress the MOT by increasing the detuning and decreasing the intensity of the cooling laser within 100 ms.
After that, a 5 ms polarization gradient cooling is applied, and the atoms are optical pumped to $|5S_{1/2}, F=1\rangle$ and loaded in the crossed ODT.
We use a single-frequency and single spatial mode fiber laser with a wavelength of 1064 nm to construct the optical dipole trap.
Two ODT laser beams are controlled by two acousto-optic modulators (AOMs), and the frequencies of two ODT beams are separated by 220 MHz to avoid the interference effects.
The left ODT laser beam (L-ODT) with 85 $\rm \mu m$ radius and the right ODT laser beam (R-ODT) with 45 $\rm \mu m$ radius are separated at angles of $30^\circ$ and $-30^\circ$ with respect to $x$-axis, respectively.
Both ODT laser beams propagate in the $x-y$ plane.
The fluctuation of ODT laser power will induce the loss of atoms.
Hence we use two 20-bit DAC boards referenced to a precise voltage source as the ODT laser power controller.
The initial power of L-ODT and R-ODT are 4.3 W and 4.5 W, respectively.
The controlling power parameters $\mathbf{p}$ generated by the Bayesian optimization script communicate with FPGA through serial port and subsequently control the ODT power controller.
After the evaporative cooling, a time-of-flight absorption imaging process is applied to measure the atom number.

\subsection{Experimental results}
\begin{figure}[htbp]
  \centering\includegraphics[width=10cm]{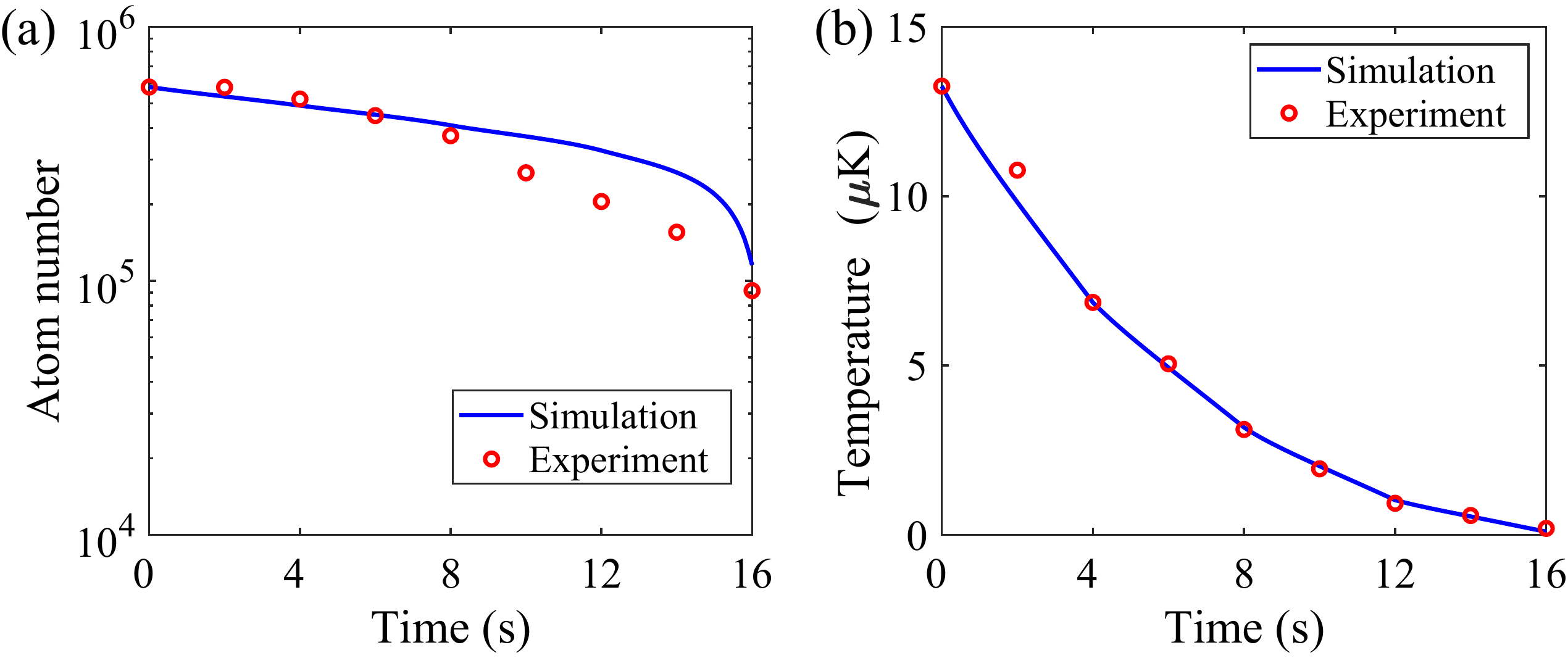}
  \caption{Comparison between the simulation results via the ECM and the experimental data. (a) Atom number versus time. (b) Temperature versus time.
  }
  \label{figure2}
\end{figure}

In this section, we experimentally demonstrate the production of BEC via Bayesian optimization based on the ECM scheme and compare them with Bayesian optimization via BBE scheme.
We first test the ECM in our experimental setup, the simulation results of atom number and temperature match the experimental data very well, as shown in Fig. \ref{figure2}. In our experiment, the total time of evaporative cooling is preset to 16 s. 

For Bayesian optimization based on the ECM scheme, the optimization score is chosen as the critical atom number $N_c$ where the phase space density $\rho_c$ equals to $2.6$.
The critical atom number $N_c$ obtained via the ECM can be considered approximately the atom number at the BEC phase transition point.
The reason why we choose the critical atom number $N_c$ as the optimization score is that the ECM is based on truncated Boltzmann distribution, and it is (is not) approximately applicable before (after) the BEC phase transition. 
Thus, we use the ECM for optimizing the atom number before the BEC phase transition point and hope to attain the optimal final atom number $N_f$ of BEC when the evaporative cooling is finished.
For Bayesian optimization via BBE scheme, the optimization score is chosen as the final atom number $N_f$ of BEC, which is obtained via time-of-flight absorption imaging at the end of evaporation. 
\begin{figure}[htbp]
  \centering\includegraphics[width=12cm]{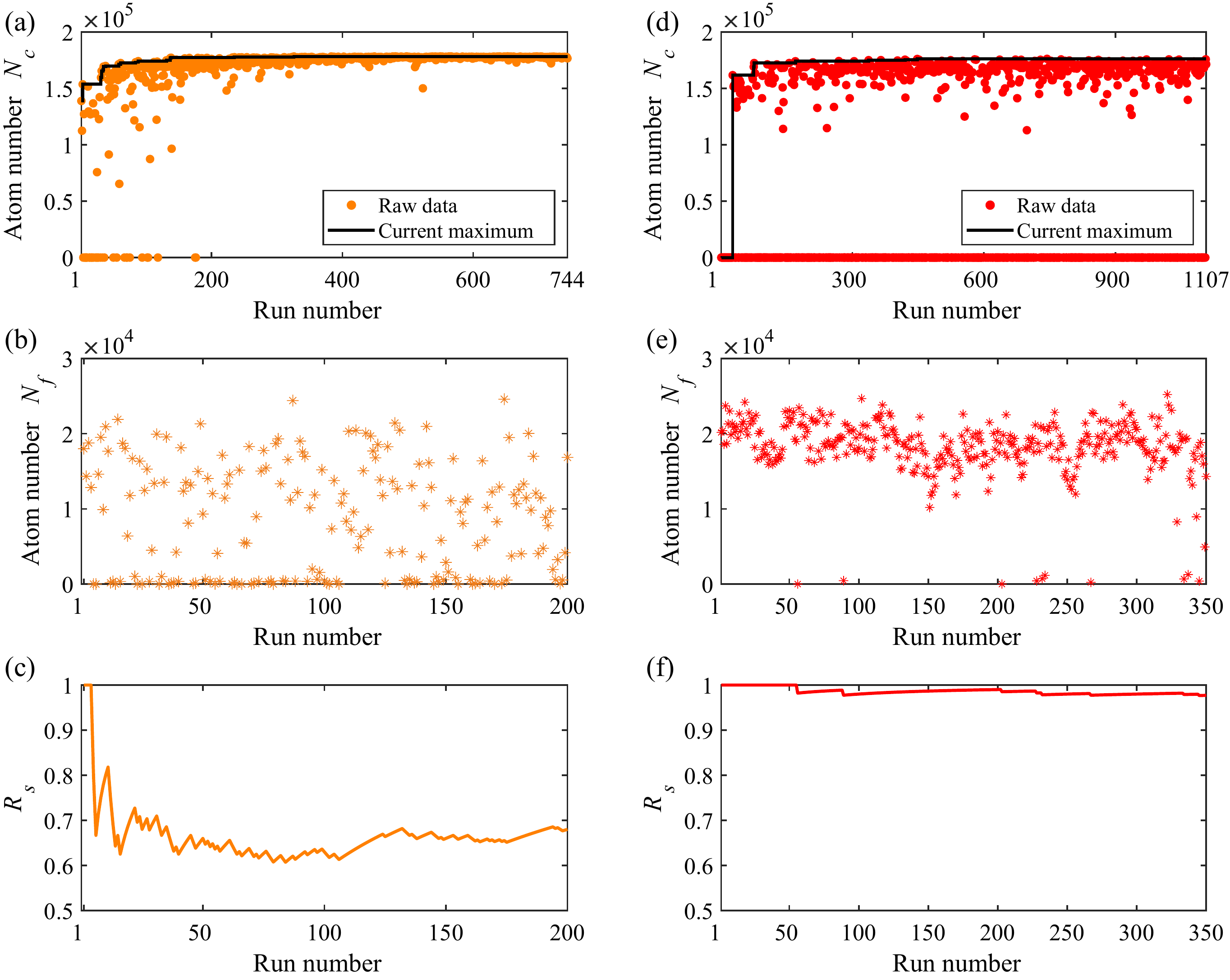} 
  \caption
  {
  Simulation and experimental results of Bayesian optimization of BEC via the ECM.
  The left-hand (right-hand) column displays results without (with) limiting the range of the final phase space density $\rho_f$.
  (a) and (d) are the simulation results of critical atom number $N_c$.
  (b) and (e) are the experimental results of the final atom number $N_f$.
  (c) and (f) are the success rate $R_s$ of creating BEC corresponding to (b) and (e), respectively. The experimental results with final atom numbers below $1 \times 10^3$ are considered failures to produce BEC.
  }
  \label{figure3}
\end{figure}

As given the initial atom number and temperature, we firstly demonstrate the simulation results of Bayesian optimization based on the ECM without limiting the final phase space density $\rho_f$ at the end of 16 s evaporative cooling, as shown in Fig. \ref{figure3}(a).
The total run number $m$ of Bayesian optimization is $744$, and then the best 200 groups are selected to implement experimentally, the result is shown in Fig. \ref{figure3}(b).
Due to the existence of experimental noise, the number of atoms extracted via absorption image below 1000 is considered a failure to create BEC.
We find that quite a few experiments fail to create BEC.
We plot the success rate $R_s$ of creating BEC in Fig. \ref{figure3}(c), where the $R_s$ means the ratio between the times of successfully getting BEC and the current total run times.

\begin{figure}[htbp]
	\centering\includegraphics[width=7.5cm]{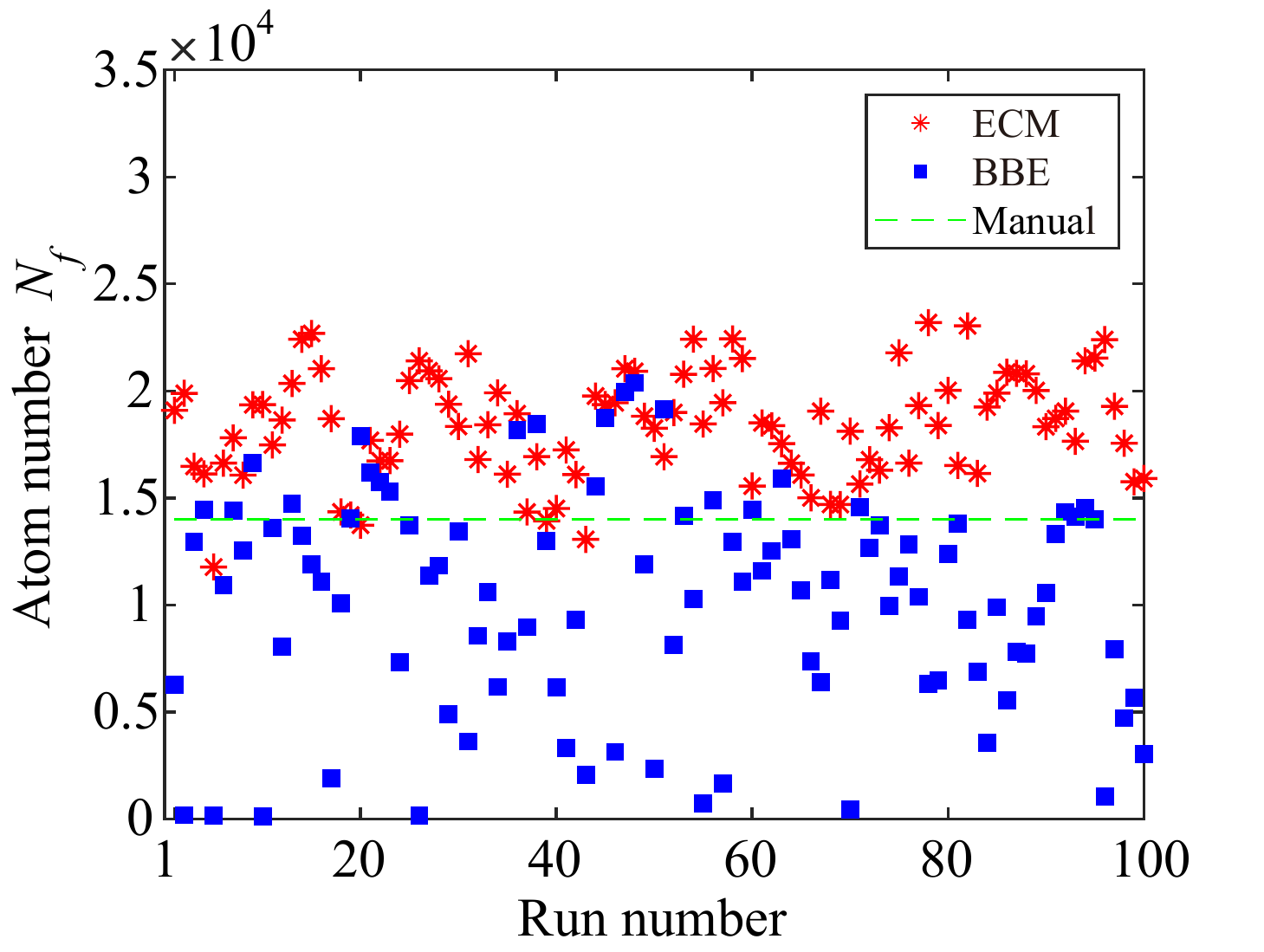} 
	\caption
	{
		Experimental results comparison of Bayesian optimization of BEC based on the ECM scheme (red asterisk) and BBE scheme (blue square).
		The green dashed line denotes the optimal atom number of pure BECs obtained via manual optimization.
	}
	\label{figure4}
\end{figure}

In order to improve the success rate of obtaining BEC, we limit the final phase space density $\rho_f$ between 30 and 200.
When the phase space density is large enough, it means that BEC is likely to occur.
Here, the optimization score $N_c$ is set as $N_c=N_c/10000$ when the Bayesian optimizer finds the final phase space density $\rho_f$ out of the range between $30$ and $200$.
Under this restriction, the Bayesian optimizer prefers to find powers of laser beams in the desired range.
The simulation results of Bayesian optimization based on the ECM with limiting the final phase space density $\rho_f$ between $30$ and $200$ are shown in Fig. \ref{figure3}(d).
The total run number $m$ of Bayesian optimization is $1107$. Then the best 350 groups are selected to implement experimentally and the experimental result is shown in Fig. \ref{figure3}(e).
The success rate $R_s$ for creating BEC is plotted in Fig. \ref{figure3}(f). We find that the success rate of obtaining BEC is greatly increased under limiting the final phase space density $\rho_f$.

\begin{figure}[hbtp]
	\centering\includegraphics[width=12cm]{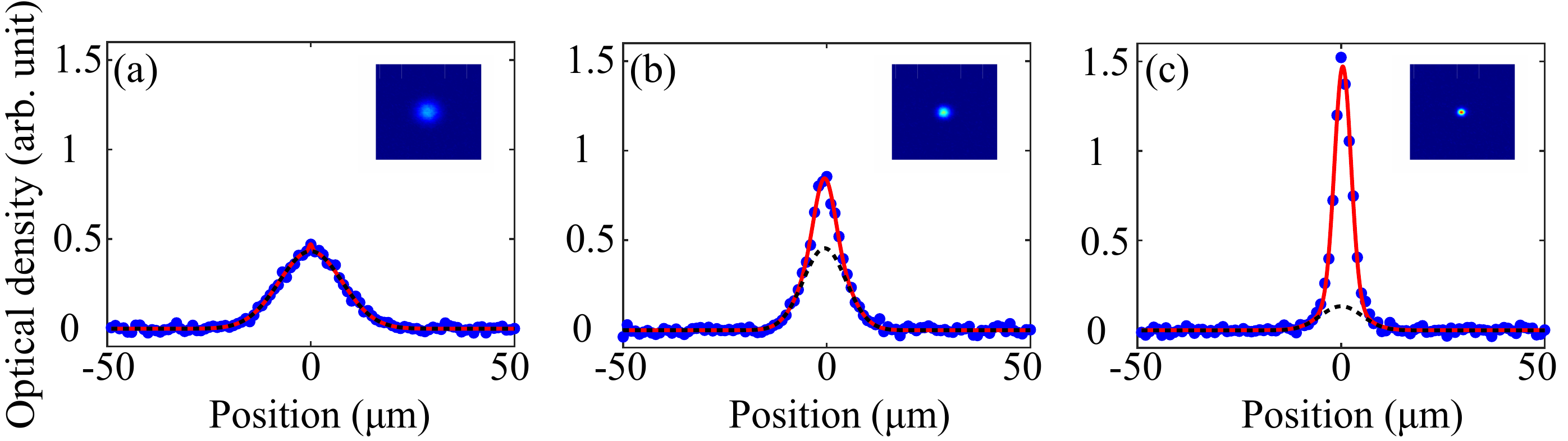} 
	\caption
	{
		Absorption images (insets) and vertically integrated optical density profiles obtained via $20$ ms time-of-flight absorption imaging for different evaporation procedures. (a) Thermal gas. (b) Mixed thermal and condensate gas. (c) Pure BEC.
		The one-dimensional cross sections (blue-filled circles) are extracted from corresponding two-dimensional absorption images along $y$-axis through the maximum. 
		The black dashed lines are Gaussian fittings of the thermal tails of atomic clouds, and the red lines are two Gaussian functions fittings of the whole atomic clouds.
		The atom number of (a), (b), and (c) are $1.1\times10^5$, $4.1\times10^4$ and $2.4\times10^4$, respectively.
	}
	\label{figure5}
\end{figure}

We compare the results of Bayesian optimization via the ECM scheme and BBE scheme, as shown in Fig. \ref{figure4}.
For Bayesian optimization based on BBE scheme, the total run number of Bayesian optimization is $100$, and the optimal atom number of pure BEC is $2.0\times10^4$.
For Bayesian optimization based on the ECM scheme, the best 100 groups are selected among $1107$ run number of Bayesian optimization to implement experimentally, and the optimal atom number of pure BEC is $2.4\times10^4$.
We obtain the optimal atom number of pure BEC with $1.4\times10^4$ via manual optimization\cite{Zhu_Ma_103701}.
The result of Bayesian optimization based on the ECM scheme performs better than that based on BBE scheme within 100 run numbers and manual optimization.
In comparison with Bayesian optimization based on blackbox experiments in which real-time experimental data are required as feedback in each cycle, our protocol does not depend on online experimental feedback in the process of optimization, but only a few experiments are required to verify some close-to-optimal curves of ODT laser powers. Therefore, this scheme can effectively save the experimental resources.
The process of BEC transition with the optimal atom number obtained via Bayesian optimization based on the ECM is shown in Fig. \ref{figure5}.

\section{Conclusion}

In this paper, we propose a Bayesian optimization of Bose-Einstein condensation via the evaporative cooling model.
In this scheme, the experimental data of evaporative cooling is replaced by the numerical simulation of the evaporative cooling model and Bayesian optimization is used for searching some close-to-optimal curves of ODT laser powers.
Among some close-to-optimal curves attained by this protocol, we experimentally demonstrate that pure BEC of $^{87}$Rb with $2.4\times10^4$ atoms can be prepared via evaporative cooling from the initial stage when the atom number is $6.0\times10^5$ at temperature 12 $\mu$K.
In comparison with Bayesian optimization based on blackbox experiment in which real-time experimental data is required as feedback in each cycle, our protocol does not depend on experiment in the process of optimization, but only a few experiments are required to verify some close-to-optimal curves of laser powers, so that this protocol can greatly save experimental resources.
\begin{backmatter}
\bmsection{Funding}
This work is supported by the National Key Research and Development Program of China (2022YFA1404104), the National Natural Science Foundation of China (12025509, 11874434, 12104521), the Key-Area Research and Development Program of GuangDong Province (2019B030330001).

\bmsection{Disclosures}
The authors declare that they have no conflict of interest.

\bmsection{Data Availability Statement}
Data underlying the results presented in this paper are not publicly available at this time but may be obtained from the authors upon reasonable request.

\end{backmatter}

\bibliography{reference}

\end{document}